\documentclass[superscriptaddress,aps,prl,floats,onecolumn,twoside,floatfix]{revtex4}
\usepackage{graphicx}
\usepackage{booktabs}
\usepackage{multirow}
\newcommand{\ba}{\begin{array}}
\newcommand{\ea}{\end{array}}
\newcommand{\be}{\begin{equation}}
\newcommand{\ee}{\end{equation}}

\begin{document}

\title{Interface Energy in the Edwards-Anderson model}  

\author{Pierluigi Contucci}
\affiliation{
Universit\`{a} di Bologna, Piazza di Porta S.Donato 5, 40127 Bologna, Italy}

\author{Cristian Giardin\`a}
\affiliation{Universit\`a di Modena e Reggio Emilia, via Allegri 9,
42100 Reggio Emilia, Italy}

\author{Claudio Giberti}
\affiliation{ Universit\`a di Modena e Reggio Emilia, via G.
Amendola 2 -Pad. Morselli- 42100 Reggio Emilia, Italy}

\author{Giorgio Parisi}
\affiliation{
Sapienza Universit\`a di Roma,  Roma, Italy, IPCF-CNR and INFN, sezione di Roma.}

\author{Cecilia Vernia}
\affiliation{
Universit\`{a} di Modena e Reggio Emilia, via Campi 213/B, 41100 Modena, Italy}

\begin{abstract}
We numerically investigate the spin glass energy interface problem in three dimensions. We analyze
the energy cost of changing the overlap from $-1$ to $+1$ at one boundary of two coupled
systems (in the other boundary the overlap is kept fixed to $+1$). We implement a parallel tempering
algorithm that simulates finite temperature systems and works with both cubic lattices and parallelepiped
with fixed aspect ratio. We find results consistent with a lower critical dimension $D_c=2.5$. The results show a good agreement
with the mean field theory predictions.
\end{abstract}

\maketitle

\section{Introduction}

It is by now a well-established
fact (numerically) that spin glass models with nearest neighbour
interactions have a finite transition temperature in spatial
dimension $D=3$ at zero magnetic field \cite{HPV}, moreover the predictions of broken replica theory are well in agreement with the numerical data \cite{JANUS}. However the value of the lower critical dimension
$D_c$, i.e. the dimension above which there is a phase transition at
strictly positive temperature, is still debated and there have been
different predictions based on different approaches. The initial
investigations, in the framework of the droplet approach, used the
{\em stiffness} exponent $\theta$ to conclude that, at zero
temperature, $\theta >0$ in $D=3$ and $\theta <0$ in $D=2$, thus
implying $2<D_c<3$ \cite{BM}. The absence of a transition in $D=2$ was confirmed also by direct numerical simulations \cite{D2}, that strongly suggest that the lower critical dimension is greater than 2.

A recent numerical
study by Boettcher \cite{B} reached the conclusion that $D_c = 2.5$
by measuring the stiffness exponent in various dimensions and making
a best fit for its dimensional dependence which gives $\theta = 0$
for $D_c = 2.5$.

An alternative approach has been put forward by Franz, Parisi,
Virasoro in \cite{FPV}. Here to measure the stability of the low temperature
spin-glass phase against thermal fluctuations, the energy cost
associated to imposing a different value of the overlap along one
spatial direction has been considered.
Remarkably \cite{FPV} founds, using an analytical
computation  based on a mean-field approximation, the same result
of the stiffness approach, i.e. $D_c = 2.5$.

However the approach introduced in \cite{FPV}
has not yet been tested on a finite-dimensional model. This is the
aim of this work. We find numerical results compatible with the prediction
$D_c = 2.5$, even though strong finite size effects prevent a final answer.

\section{Theoretical prediction}

\begin{figure}
\setlength{\unitlength}{1cm}
          \centering
                \includegraphics[width=11cm,height=9cm]{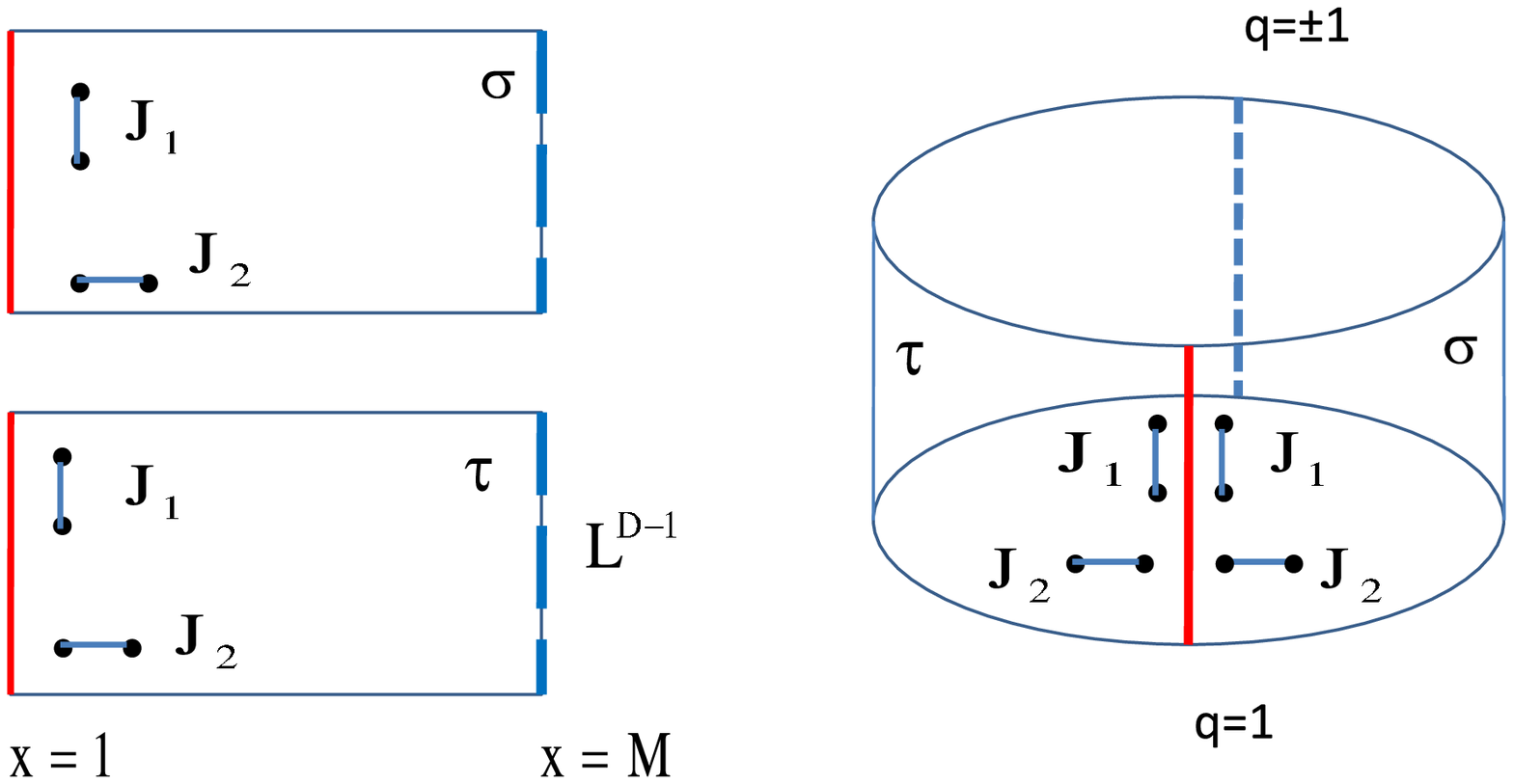}
                \caption{From two copies of longitudinal size $M$ to one system
                of longitudinal size $2(M-1)$.}
\label{Fig1}
\end{figure}

We first recall the approach of \cite{FPV}.
We consider {\em two} identical copies (same disorder) of a system
constrained to have their mutual overlaps on the boundaries fixed to
some preassigned values. More precisely, we take two systems with the
parallelepiped geometry of Fig. (\ref{Fig1}). Each system has
a base area $L^{D-1}$ and height $M$;  we denote by $x$ the coordinate
along the longitudinal direction of length $M$. The study of systems with different
aspect ratio $R$ for determining the interface exponent has been used in the past \cite{VARI}.
Each of the two systems has free boundary conditions along the longitudinal
direction and periodic boundary conditions along the
transverse direction. However the two systems are
coupled by the value of the overlap at their boundary
planes. We compare the following two situations.
In the first case, which we call $(++)$ coupling condition,
we impose an overlap $+1$ on both the left
$(x=1)$ and the right $(x=M)$ boundary plane; in the second case, called here the $(+-)$
coupling condition, the overlap on the left boundary $(x=1)$ is
equal to $+1$ and the overlap on the right boundary $(x=M)$ is
equal to $-1$. The quantity that we will be interested in is the
difference of free energies between the two cases
\begin{equation}
\label{deltaF}
\Delta F = - \frac{1}{\beta} Av(\log Z_{+-} - \log Z_{++})
\end{equation}
where $\beta$ is the inverse critical temperature,
$Av(-)$ denotes the average over the random couplings and
\begin{equation}
Z_{ab} = \sum_{\sigma\tau}  e^{-\beta (H_{\sigma}+H_{\tau})}
\delta\left(\frac{1}{L^{D-1}}\sum_{i \in A_{1}} {\sigma_{i}\tau_i -a}\right)
\delta\left(\frac{1}{L^{D-1}}\sum_{i \in A_{M}} {\sigma_{i}\tau_i -b}\right)
\end{equation}
with $a,b\in \{\pm \}$ and $A_{1}$ (resp. $A_{M}$)
the left (resp. right) boundary plane, $H_{\sigma}$
the Hamiltonian of the spin configuration $\sigma$.
One can prove that $\Delta F$ is a positive quantity thanks to the fact
that $Z_{+,+}\ge Z_{+,-}$. Thus Eq. (\ref{deltaF}) seems an
appropriate definition of a domain wall in spin glass models,
in full analogy with usual ferromagnetic systems where a
positive domain wall is obtained by imposing different
values of the magnetization along one spatial direction.

Since free energies are not directly accessible through
Monte Carlo methods \cite{BM,BFP}, in our simulations we measured the
differences of internal energies between the $(+-)$ and $(++)$
case
\begin{equation}
\Delta E = Av(E_{+-} - E_{++})
\end{equation}
where
\begin{equation}
E_{ab} = \sum_{\sigma\tau} (H_{\sigma}+H_{\tau}) \frac{e^{-\beta (H_{\sigma}+H_{\tau})}
\delta\left(\frac{1}{L^{D-1}}\sum_{i\in A_1} {\sigma_{i}\tau_i -a}\right)
\delta\left(\frac{1}{L^{D-1}}\sum_{i\in A_{M}} {\sigma_{i}\tau_i -b}\right)}{Z_{ab}}.
\end{equation}

\noindent
We observe that at zero temperature $\Delta E = \Delta F$.
Thus, at small enough temperatures, the positivity of $\Delta E$
follows from the positivity of $\Delta F$ by a continuity argument.
However we remark that in our numerical simulations
we found {\em always}  positive energy differences
at all temperatures. More precisely, for a random
initial condition, the sign of $E_{+-} - E_{++}$ fluctuates
at the beginning of the Monte Carlo run and then it
stabilizes to a positive sign.

\vspace{0.5cm}

For many models \cite{FPV} we observe  that, below the critical temperature, the free energy cost,
as a function of the transverse and longitudinal length,
scales for large volumes as
$
\Delta F \sim L^{D-1} M^{-\omega}
$
with $\omega$ independent from the dimension (for $T>T_c$,
$\Delta F \sim \exp(-L)$).
The scaling of the free energy cost defines a lower
critical dimension $D_c$ in the usual way: for an assigned aspect ratio
$R = M/L$ the free energy cost behaves as
$\Delta F \sim L^{D-1-\omega}$. From this expression we see that
the lower critical dimension, the one above which $\Delta F$ increases
with $L$ and below which it vanishes with $L$,
turns out to be $D_c = \omega + 1$.
The mean-field computation of \cite{FPV} predicts
$D_c = 5/2$, i.e. $\Delta F \sim L^{D-1} M^{-3/2}$.
In other words, below the critical temperature, the free energy cost divided by the volume $V$ turns
out to be independent from $L$ and, for large $L$, it scales as
\begin{equation}
\beta \Delta F/V \approx M^{-5/2}f(\beta).
\end{equation}
Since the free energy difference and the internal energy difference scale
with the same exponent below the critical temperature then we also have
\begin{equation}
\label{deltaEsottoTc}
\Delta E/V \approx M^{-5/2}f'(\beta).
\end{equation}


Generally speaking, at the phase transition point one usually obtains
by changing the boundary conditions that
$$
\Delta F \to C \ne 0.
$$
Using Widom's scaling one gets:
\begin{equation}
\label{deltaEaTc}
\frac{\Delta E}{V} \sim M^{-3 + 1/\nu}.
\end{equation}
In a given case there could be cancellations and $\Delta F$ at the
critical point could go to zero with the volume if $C=0$. In ref. \cite{BFP}
it was shown by an explicit computation that this does not happen, at
least in the mean field case.  We expect that $C$ is a smooth function
of the dimension, so that $C=0$ in $D=3$, would be surprising.

\section{Numerical implementation}

In our simulations we have considered the above mentioned geometrical
set-up for the case of the Edwards-Anderson Hamiltonian with dichotomic ($\pm 1$)  symmetrically
distributed random couplings. In particular, the implementation of the two
systems coupled at their boundaries was obtained considering a parallelepiped with
{\em doubled size in the longitudinal direction}.
Denoting by $y$ the coordinate in the longitudinal direction
of the doubled parallelepiped, the disorder variables are symmetric with
respect to the central plane $y=M$. The sub-systems in the half-volumes $y<M$ and $y>M$
interact in the following way (see Fig (\ref{Fig1})):

i) to keep the overlap of the spins $\sigma$ and $\tau$ on the old $x=1$ plane fixed to $+1$,
the spins of the new doubled system are identified and they lie on the plane $y=M$;

ii) to keep the overlap of the spins $\sigma$ and $\tau$ on the old $x=M$ plane fixed to either $\pm1$,
the spins of the new doubled system at position $y=1$ and $y=2M-1$ are identified.
The strength of the couplings among the spins inside those planes are
reduced by a factor one-half (in order to have un-biased interactions);
the couplings between nearest neighbors not lying inside those planes
are flipped when going from the $(++)$ case to the $(+-)$.

Note that the effective longitudinal size of the total system is $2(M-1)$.

\section{Results}

The data that we are going to describe concern systems with sizes $4\le L \le 10$ for the cubic lattice
whereas $4\le L \le 8$ and $4 \le M \le 25$ for the parallelepiped lattice (see Tab. 1)
and with $30$ different values of the aspect ratio $R=M/L$, ranging from $0.57$ up to $6.25$.
The computations have been performed implementing a multi-spin coding of the parallel tempering algorithm
with 33 equally spaced temperatures ranging between 0.6 and 2.2.
\begin{table}
\begin{center}
\begin{tabular}{|c|c|c|c|}\hline
L &M & sweeps & sample\\ \hline
\multirow{8}*{4} & 4 & 8192 & 9600\\
\cline{2-4}
&  5 & 8192 & 6400\\
\cline{2-4}
&  7 & 8192 & 6400\\
\cline{2-4}
&  9 & 32768 & 6400\\
\cline{2-4}
&  13 & 262144 & 3200\\
\cline{2-4}
&  17 & 1048576 & 3200\\
\cline{2-4}
&  21 & 4194304 & 3200\\
\cline{2-4}
&  25 & 4194304 & 2560\\
\hline
\end{tabular}
\hspace{0.5cm}
\begin{tabular}{|c|c|c|c|}\hline
L & M & sweeps & sample\\ \hline
\multirow{9}*{6} & 4 & 8192 & 6400\\
\cline{2-4}
&  5 & 8192 & 9600\\
\cline{2-4}
&  6 & 8192 & 6400\\
\cline{2-4}
&  7 & 8192 & 6400\\
\cline{2-4}
&  9 & 8192 & 6400\\
\cline{2-4}
&  13 & 32768 & 3200\\
\cline{2-4}
&  17 & 131072 & 1920\\
\cline{2-4}
&  21 & 1048576 & 1920\\
\cline{2-4}
&  25 & 2097152 & 1280\\
\hline
\end{tabular}
\hspace{0.5cm}
\begin{tabular}{|c|c|c|c|}\hline
L & M & sweeps & sample\\ \hline
\multirow{8}*{7} & 4 & 8192 & 6400\\
\cline{2-4}
&  5 & 8192 & 6400\\
\cline{2-4}
&  7\ & 8192 & 6400\\
\cline{2-4}
&  9 & 8192 & 6400\\
\cline{2-4}
&  13 & 65536 & 6400\\
\cline{2-4}
&  17 & 524288 & 3200\\
\cline{2-4}
&  21 & 8388608 & 1280\\
\cline{2-4}
&  25 & 8388608 & 1280\\
\hline
\end{tabular}
\hspace{0.5cm}
\begin{tabular}{|c|c|c|c|}\hline
L & M & sweeps & sample\\ \hline
\multirow{5}*{8} & 5 & 8192 & 6400\\
\cline{2-4}
&  7 & 8192 & 6400\\
\cline{2-4}
&  9 & 8192 & 6400\\
\cline{2-4}
&  13 & 65536 & 3200\\
\cline{2-4}
&  17 & 524288 & 1920\\
\hline
\end{tabular}
\end{center}
\caption{System size parameters for the numerical simulations.}
\end{table}

\begin{figure}
\setlength{\unitlength}{1cm}
          \centering
                \includegraphics[width=9cm,height=9cm]{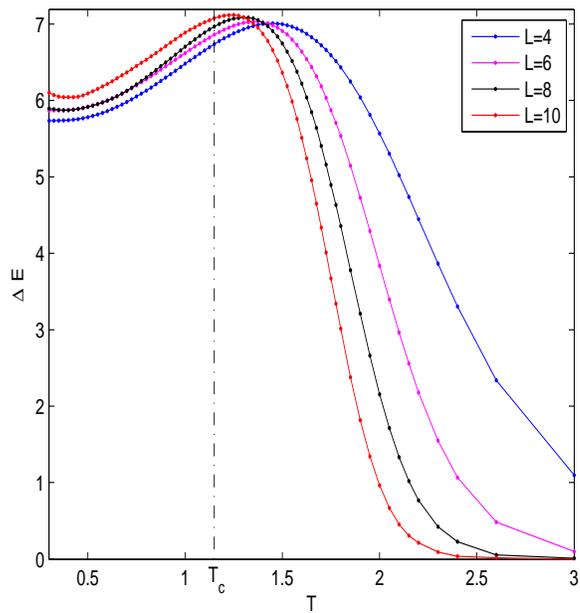}
                \caption{Cubic lattice: extensive internal energy cost of an interface versus the temperature.}
\label{Fig2}
\end{figure}

We start from the analysis of a cubic lattice ($M=L$).
The data for the extensive internal energy difference
as a function of temperature for different system sizes
are shown in Fig. (\ref{Fig2}).
Below the critical temperature $T_c \approx 1.15$
the data vary weakly with the linear lattice size $L$.
The theoretical prediction would have been $\Delta F \sim L^{0.5}$.
This suggested that there might be strong finite size effects
on the energy interface cost for the small lattices we consider.

\begin{figure}
\setlength{\unitlength}{1cm}
\begin{minipage}[h!]{7cm}
          \centering
               \includegraphics[width=7cm,height=7cm]{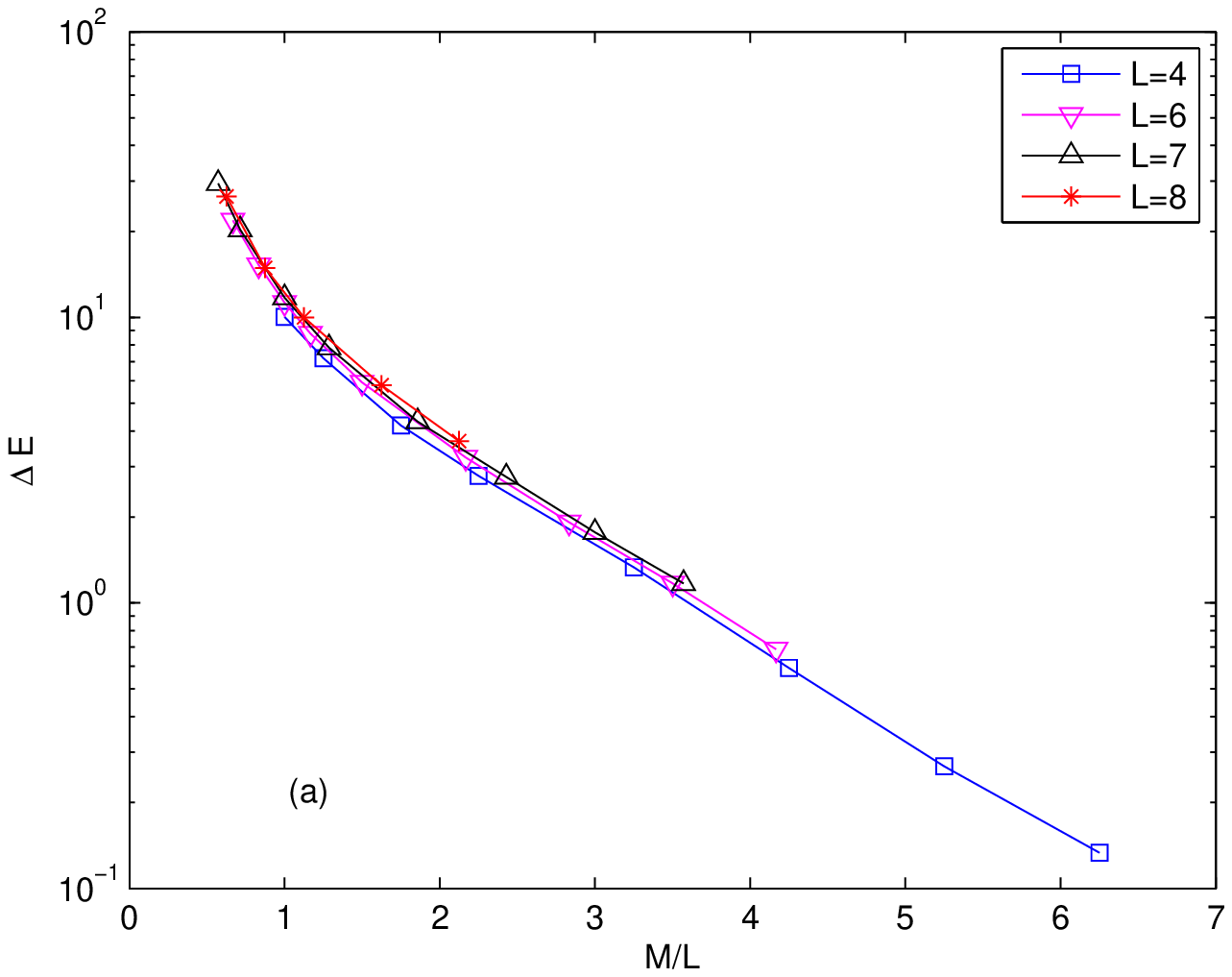}
\end{minipage}
\ \hspace{.3cm} \
\begin{minipage}[h!]{7cm}
\centering
               \includegraphics[width=7cm,height=7cm]{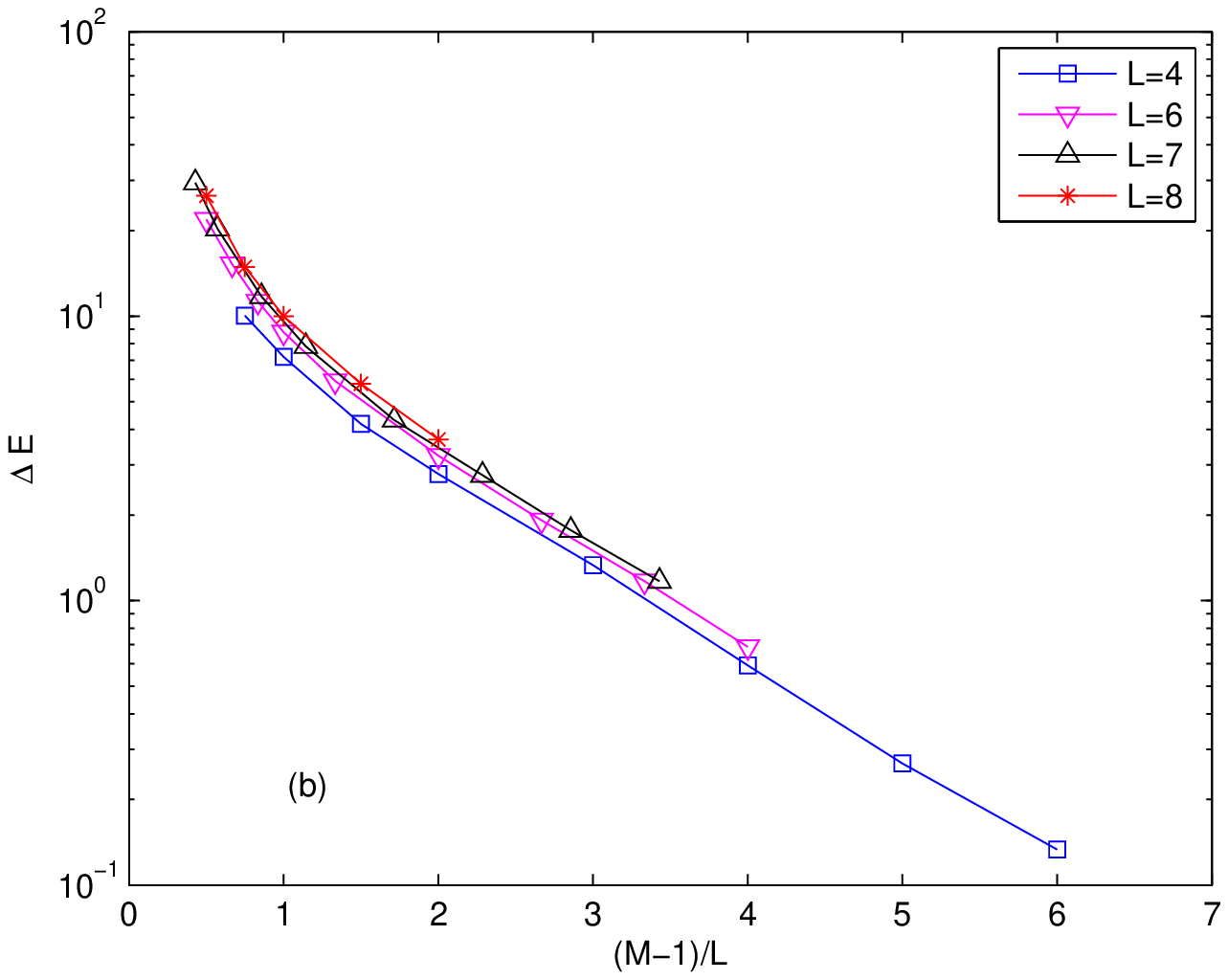}
\end{minipage}

\setlength{\unitlength}{1cm}
\begin{minipage}[h!]{7cm}
          \centering
               \includegraphics[width=7cm,height=7cm]{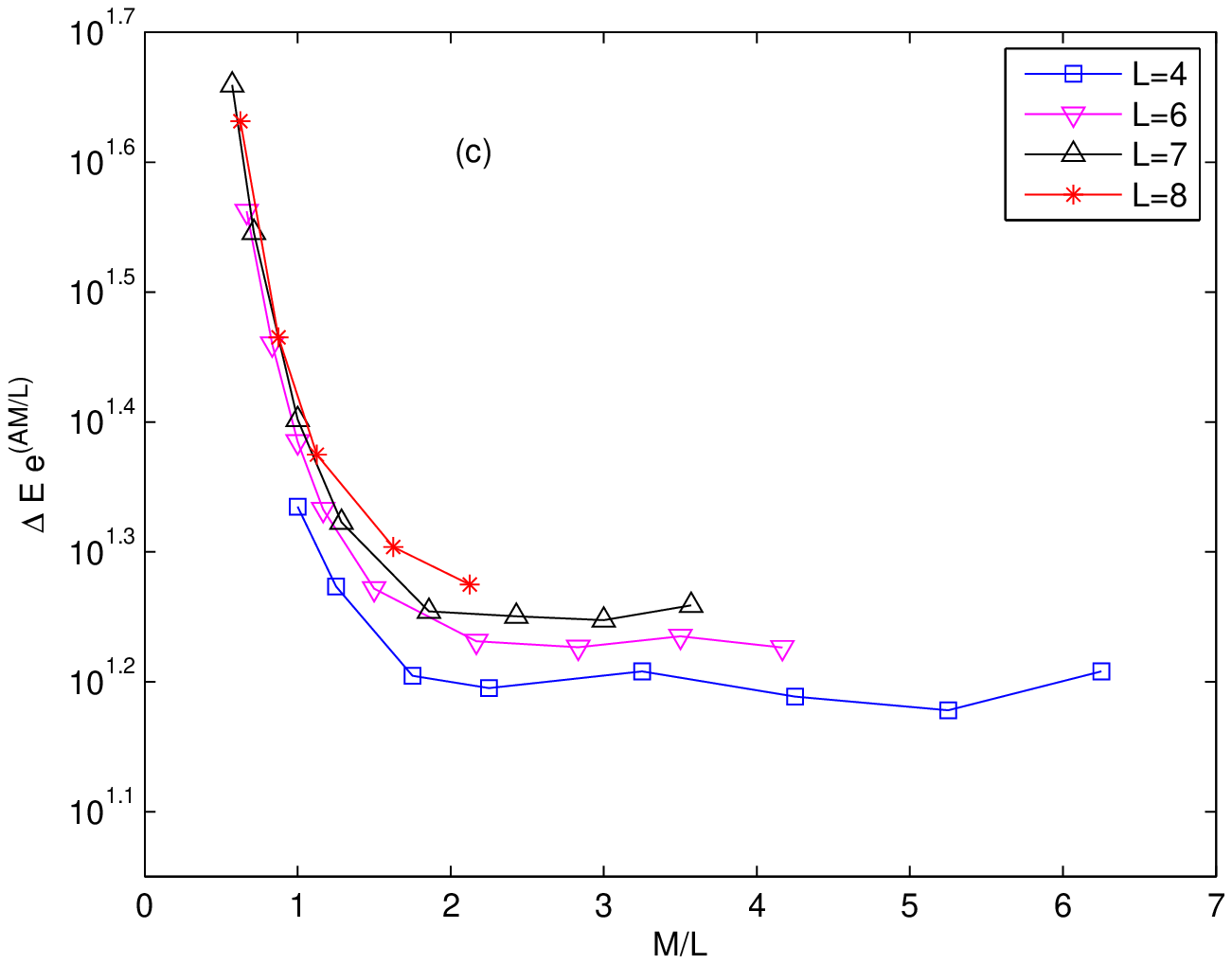}
\end{minipage}
\ \hspace{.3cm} \
\begin{minipage}[h!]{7cm}
\centering
               \includegraphics[width=7cm,height=7cm]{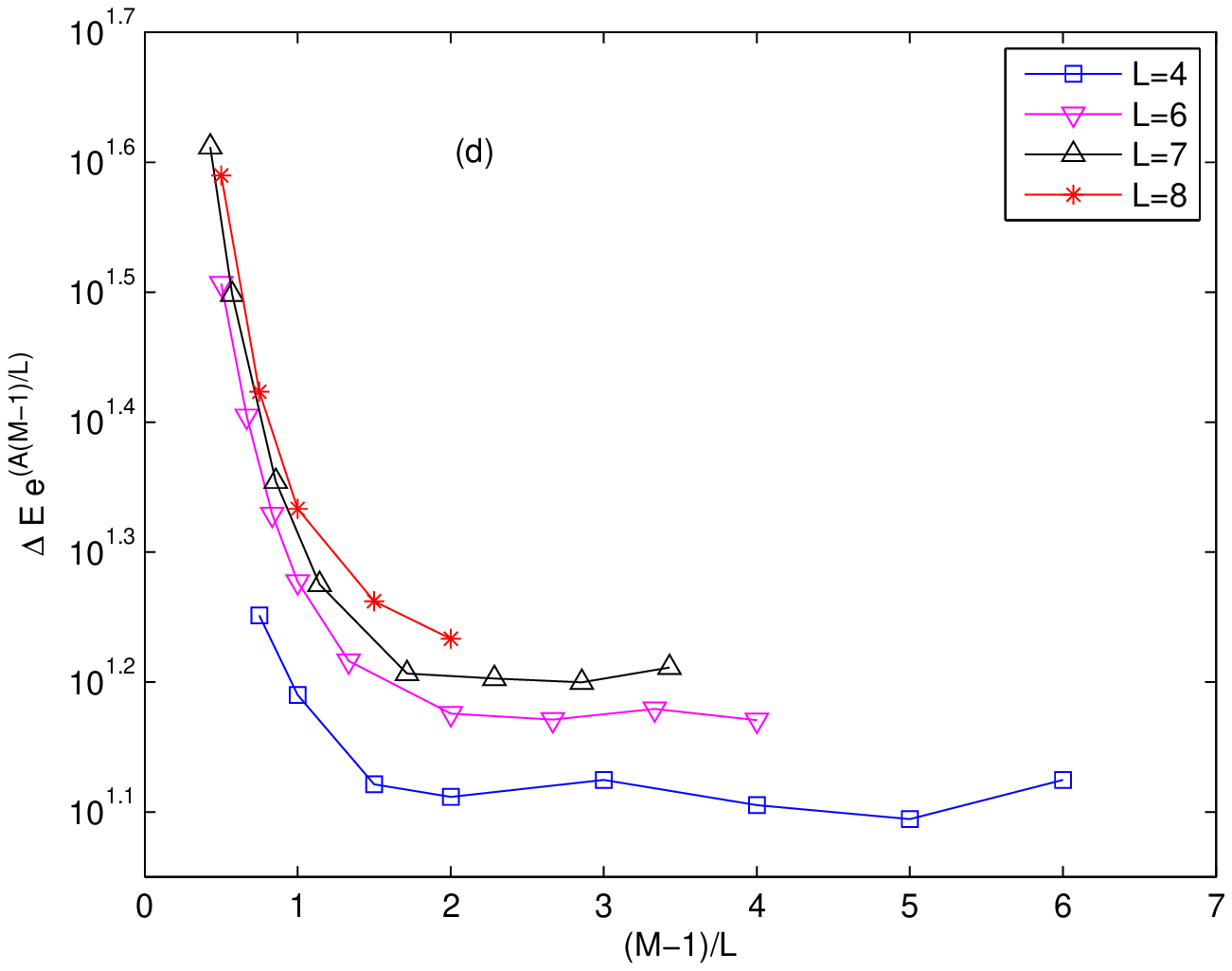}
\end{minipage}

              \caption{$\Delta E$ as a function of $M/L$ (panel (a)) and
              as a function of $(M-1)/L$ (panel (b)) at temperature $T=1.1$;
              $\Delta E e^{A M/L}$ as a function of  $M/L$ (panel (c)) and
              $\Delta E e^{A (M-1)/L}$ as a function of $(M-1)/L$ (panel (d)).
              $A$ is $0.77$. In each panel the different curves
              refer to different $L$ values.}
\label{Fig3}
\end{figure}

\vspace{0.5 cm}
This is confirmed by the analysis around the critical temperature
which is displayed in Fig. (\ref{Fig3}). For the parallelepiped geometry,
the extensive internal energy interface cost shows a different
behavior when plotted versus the aspect ratio
$R = M/L$ or against the ``corrected''
aspect ration $\tilde R = (M-1)/L$ (see the panels (a) and (b) of
Fig. (\ref{Fig3})). Even though these two characterizations of the
aspect ratios are totally equivalent in the thermodynamic limit,
the data show a marked difference in the two cases.
To better appreciate this difference we show in
the lower panels (c) and (d) of Fig. (\ref{Fig3}) the same data multiplied by
$e^{A R}$ on the left and $e^{A\tilde R}$ on the right,
where $A=0.77$ has been computed fitting to an exponential
form the behavior of the data at large $R$ (or $\tilde R$,
it gives the same $A$).
We do not claim an exponential dependence, we merely used
the exponential re-scaling from the fit to make
the data variation more visible.

\vspace{0.5 cm}
To extract sensible information from our data, a careful analysis of the
dependence of $\Delta E$ from $L$ and $M$ is required.
Fig. (\ref{Fig3}) tell us that it is difficult to
make an analysis of the data as a function of the aspect ratio \cite{CBM}.
Thus we start from the expected behavior of very large systems
$$
\Delta E \sim L^2 f(M)
$$
which is suggested by the general argument that each of the $L^2$ spins
on each plane give the same contribution.
Thus, for the small finite systems we have, we study
the dependence on $1/L$ for a fixed $M$.
Since we want to study the behavior for systems
with longitudinal size much smaller than
transverse size, we must limit the analysis
to small values of $M$.

\begin{figure}[h]

\setlength{\unitlength}{1cm}
\centering
               \includegraphics[width=8cm,height=8cm]{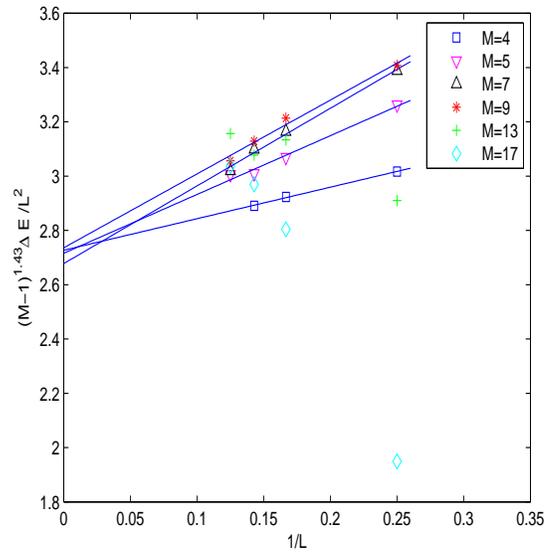}

\caption{$\frac{\Delta E}{L^2} (M-1)^{1.43}$ versus $1/L$ at temperature $T=T_c$.}
\label{Fig4}

\end{figure}

At critical temperature we expect the scaling of Eq. (\ref{deltaEaTc}),
i.e.
$$
\frac{\Delta E}{L^2} \sim M^{-2 + 1/\nu}
$$
Figure (\ref{Fig4}) shows that $\frac{\Delta E}{L^2} (M-1)^{1.43}$
does not depend on $L$ for large $L$ values.
This implies a value of $1/\nu = 0.57$
to be compared to $1/\nu = 0.408$ of \cite{HPV}.
The predicted value of our exponent should be $1.75$. We think that the
difference between $2.45$ and the predicted value ($1.75$) is not
worrisome given the fact that the value of $\nu$ was computed in  \cite{HPV} using
lattices (up to $40^3$) that are much larger of ours and, historically,
previous determinations of $1/\nu$ on smaller lattices gave large values
of $1/\nu$.

\begin{figure}[h]
    \setlength{\unitlength}{1cm}
          \centering
               \includegraphics[width=8cm,height=8cm]{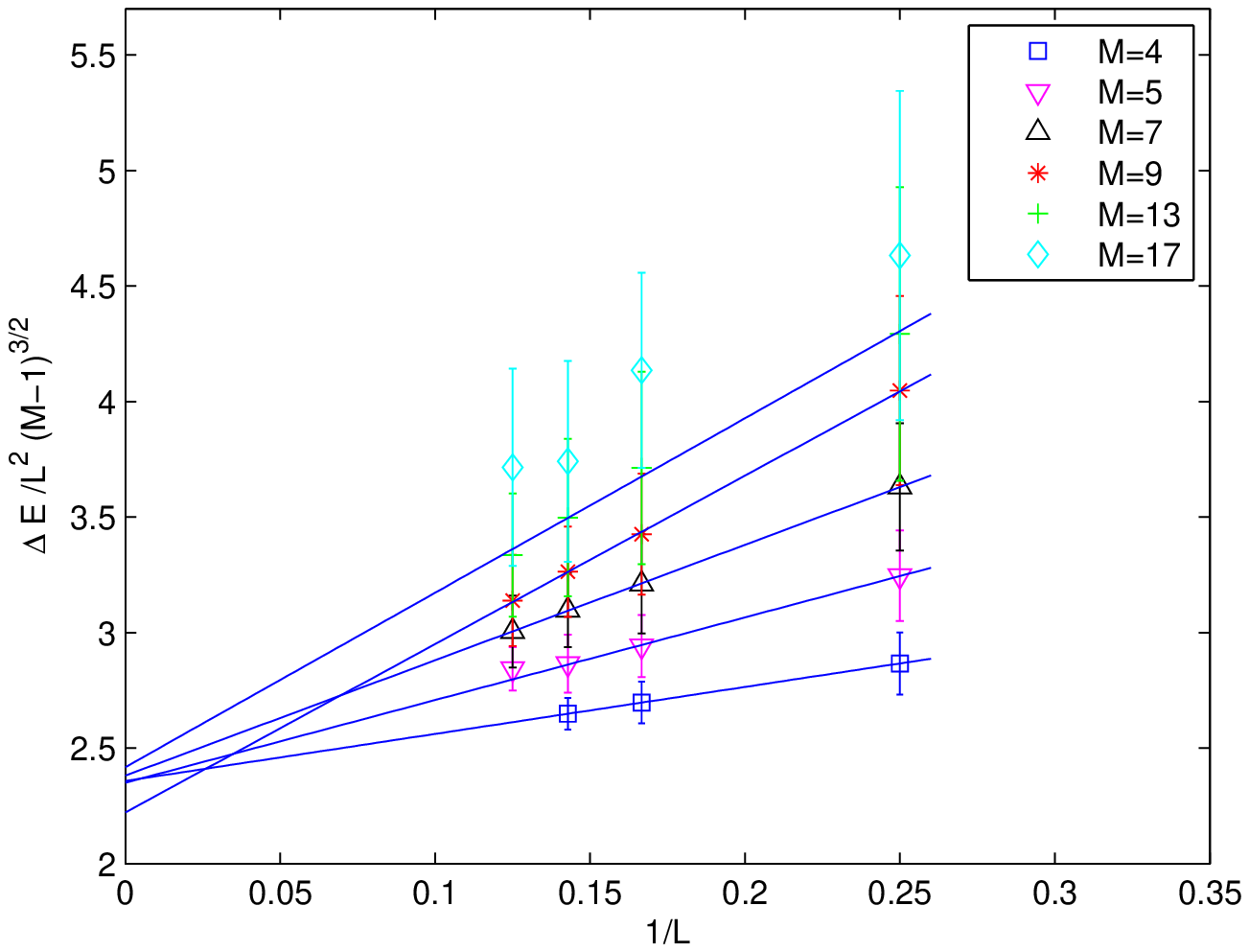}
               \caption{$\frac{\Delta E}{L^2} (M-1)^{3/2}$ versus $1/L$ at temperature $T=0.6$.}
               \label{Fig5}
\end{figure}

\begin{figure}[h]
    \setlength{\unitlength}{1cm}
               \includegraphics[width=8cm,height=8cm]{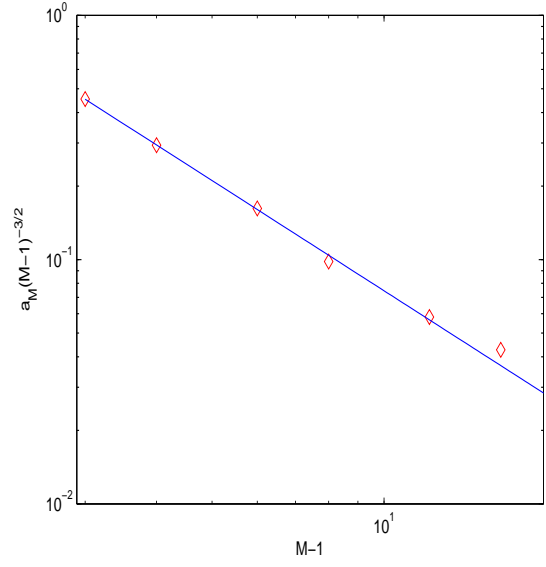}
               \caption{The intercepts of Fig. (\ref{Fig5}) as a function of $M-1$ at temperature $T=0.6$
               in log-log scale. The straight line is the best fit of the form $a (M-1)^{-3/2}$.}
               \label{Fig6}
\end{figure}

\vspace{0.5 cm}
Below the critical temperature we expect the scaling of Eq. (\ref{deltaEsottoTc}),
i.e.
$$
\frac{\Delta E}{L^2} \sim M^{-3/2}
$$
In Fig. (\ref{Fig5}) we show $\frac{\Delta E}{L^2} (M-1)^{3/2}$ versus $1/L$.
The behavior is described by a fit of the form $a_M + \frac{b_M}{L}$.
The slope $b_M$ is increasing with $M$ while the intercept $a_M$ is instead
constant in $M$.
In Fig. (\ref{Fig6}) we show  $a_M (M-1)^{-3/2}$ versus $M-1$ in log-log scale.
The data nicely fall on a straight line with a constant slope $a$.
The best fit on $b_M$ gives $k(M-1)^{-\delta} + h$
with $k<0$ and $\delta=-0.22$ if $M\in \{4,5,7,9,13\}$ while
$\delta=-1/2$ excluding the values $M\ge 9$. Thus it would imply
that the correction $\frac{b_M}{L}$ converges to zero for large
volumes with an assigned aspect ratio.

\vspace{0.5cm}
Summarizing, we have found that in the regime $M << L$ the data
for $\Delta E$ are well described by the following behavior
$$
\Delta E \sim L^2 (M-1)^{-3/2} \left(a + \frac{b_M}{L}\right)
$$
which is compatible with the prediction $D_c = 2.5$.

\begin{figure}
\setlength{\unitlength}{1cm}
          \centering
               \includegraphics[width=8cm,height=8cm]{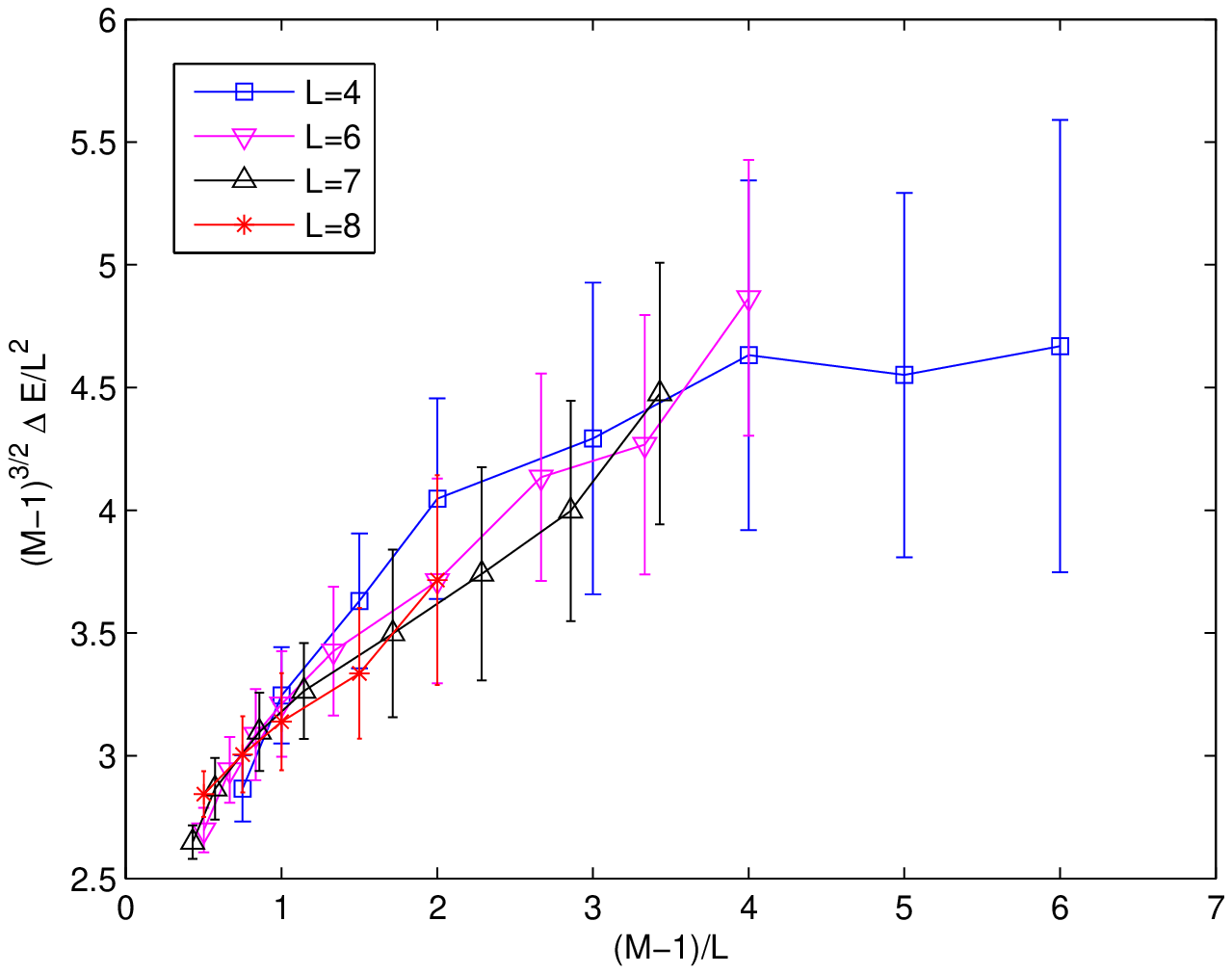}
               \caption{$\frac{\Delta E}{L^2}(M-1)^{3/2}$ versus $(M-1)/L$ at temperature $T=0.6$.}
\label{Fig7}
\end{figure}

It would nice to study the scaling property of the interface
energy as a function of both variables $L,M$.
In particular one would like to study the dependence
when both $L$ and $M$ grow with a fixed aspect ratio.
In Fig. (\ref{Fig7}) we plot
$\frac{\Delta E}{L^2}(M-1)^{3/2}$ versus $(M-1)/L$ at temperature $T=0.6$.
We see that there is not a global clear scaling of the data.
We tried also the aspect ratio scaling in the spirit
of \cite{CBM}, according to which $\Delta E = c_0(R) L^{\theta_0(R)}$.
We found very large oscillations in the exponent $\theta_0$.
We tried to cure the finite size effects by considering
the scaling $\Delta E = \left(c_0(R) + \frac{c_1(R)}{L}\right) L^{\theta_1(R)}$
but we have too few points and the final errors have a too large error.
One would need larger lattices and smaller errors on the data
to reach a definite conclusion.

At this stage it is difficult to assign a precise error on $\omega$
because of the limited range of $L$ ($4-10$); for such small lattices it
would be mandatory to consider corrections to scaling and the fits
would become unstable.

\vspace{0.5cm}
Acknowledgments: we thank E.Brezin and S.Franz for interesting
discussions.

\newpage

\end{document}